\documentclass[prb,twocolumn,showpacs]{revtex4}
\usepackage{graphicx,epsf}
\usepackage{bm,bbm}      % bold math

\newcommand{\BEDT}{$\alpha-$(BEDT-TTF)$_2$I$_3$}

\newcommand{\bone}{\mathbbm{1}}

\newcommand{\Pib}{\mbox{\boldmath $\Pi $}}

\newcommand{\deltab}{\mbox{\boldmath $\delta $}}

\newcommand{\Hmath}{\mathcal{H}}

\newcommand{\bq}{{\bf q}}
\newcommand{\bk}{{\bf k}}
\newcommand{\br}{{\bf r}}
\newcommand{\be}{{\bf e}}
\newcommand{\ba}{{\bf a}}
\newcommand{\bv}{{\bf v}}
\newcommand{\bw}{{\bf w}}

\newcommand{\bK}{{\bf K}}
\newcommand{\bD}{{\bf D}}
\newcommand{\bA}{{\bf A}}
\newcommand{\bV}{{\bf V}}

\newcommand{\qtilde}{\tilde{q}}
\newcommand{\wtilde}{\tilde{w}}

\newcommand{\beq}{\begin{equation}}
\newcommand{\beqn}{\begin{eqnarray}}
\newcommand{\eeq}{\end{equation}}
\newcommand{\eeqn}{\end{eqnarray}}
\newcommand{\nn}{\nonumber}

\begin{document}

\def\tende#1{\,\vtop{\ialign{##\crcr\rightarrowfill\crcr
\noalign{\kern-1pt\nointerlineskip}
\hskip3.pt${\scriptstyle #1}$\hskip3.pt\crcr}}\,}

\title{Tilted anisotropic Dirac cones in quinoid-type graphene and \BEDT}
\author{M.O. Goerbig, J.-N. Fuchs, G. Montambaux, and F. Pi\'echon}

\affiliation{
Laboratoire de Physique des Solides, CNRS UMR 8502, Univ. Paris-Sud, F-91405 Orsay cedex, France}

\begin{abstract}

We investigate a generalized two-dimensional Weyl Hamiltonian, which may describe the low-energy
properties of mechanically deformed graphene and of the organic compound \BEDT~under pressure. The
associated dispersion has generically the form of tilted anisotropic Dirac cones. The tilt arises
due to next-nearest-neighbor hopping when the Dirac points, where the valence band touches the
conduction band, do not coincide with crystallographic high-symmetry points within the first
Brillouin zone. Within a semiclassical
treatment, we describe the formation of Landau levels in a strong magnetic field, the relativistic
form of which is reminiscent to that of graphene, with a renormalized Fermi velocity due to the
tilt of the Dirac cones. These relativistic Landau levels, experimentally accessible via spectroscopy
or even a quantum Hall effect measurement, may be used as a direct experimental verification of
Dirac cones in \BEDT.

\end{abstract}
\maketitle

\section{Introduction}

The discovery of a particular quantum Hall effect in graphene\cite{novoselov,zhang}
has shown that the low-energy electronic properties in this two-dimensional (2D) carbon crystal are described not in terms of a Schr\"odinger-type wave equation but
by a relativistic Dirac equation.\cite{antonioRev} Due to a $\pi$-band,
which shrinks at half-filling to two inequivalent points at the corners of the
first Brillouin zone (BZ), the electronic energy dispersion is almost linear resulting
in Dirac cones.
This is reminiscent of the case of massless relativistic particles, where the speed
of light $c$ is replaced by a Fermi velocity $v_F$, which is roughly 300 times smaller
than $c$.

Another material where Dirac cones are expected to occur is the organic 2D
compound \BEDT~under pressure.\cite{katayama,kobayashi,fukuyama} The relativistic
behavior of the carriers may be at the origin\cite{katayama} of an experimentally
observed $T^2$ dependence of the carrier density.\cite{kajita,tajima}
Whereas in graphene, the Dirac cones at the corners of the first BZ are
isotropic, they are situated within the first BZ in \BEDT, strongly anisotropic,
and {\sl tilted} in the wave-vector energy space ($\bk,E$).\cite{katayama,kobayashi}
The electronic properties are described by a generalized Weyl Hamiltonian with
terms linear in the 2D wave vector $\bk$. However, in contrast to graphene, there
is yet no direct experimental evidence for the presence of Dirac cones in \BEDT~or
whether the system is simply a narrow-gap semiconductor.

In the present paper, we study the structure of the generalized Weyl
Hamiltonian, which yields energy dispersions in form of tilted
anisotropic Dirac cones. In the presence of a strong magnetic field,
the dispersion is quantized in relativistic Landau levels (LLs),
with the characteristic $\pm \sqrt{nB}$ behavior known from
graphene. The tilt and the anisotropy of the Dirac cones give rise
to a renormalization of the effective Fermi velocity and therefore
of the typical LL spacing.

One example of a 2D system described by such generalized Weyl equation may
be the above-mentioned organic material \BEDT. We show, within an effective
tight-binding model on an anisotropic triangular lattice with two atoms per unit
cell,\cite{hotta} that the tilting of
the Dirac cones is due to next-nearest-neighbor ({\sl nnn}) hopping, which may be
in \BEDT~on the same order of magnitude as nearest-neighbor ({\sl nn}) hopping.\cite{mori,kondo}
A necessary condition for {\sl nnn} hopping to cause a tilt of the Dirac cones is that
they are situated at points in the first BZ different from those of high crystallographic symmetry,
such as its corners. Furthermore, we show that it may
equally apply to graphene
when the Dirac points, $\bD$ and $\bD'$ move away from the high-symmetry
points $\bK$ and $\bK'$ at the corners of the first BZ. In this
case the wave-vector expansion of the {\sl nnn} term yields a linear contribution,
whereas it is quadratic when the
Dirac points coincide with the BZ corners $\bK$ and $\bK'$. Such motion of the
Dirac points may indeed be induced by a quinoid-type lattice distortion\cite{pauling} of the graphene sheet. However, we show that the tilt of the Dirac cones is much
less pronounced than in \BEDT. Alternatively, this motion of Dirac points
may be studied in cold atoms in an optical
lattice where one may deform the honeycomb lattice and fine-tune the {\sl nn} and {\sl nnn}
hopping parameters with the help of the laser intensities, wavelengths, and relative
orientation.\cite{zhu}

The paper is organized as follows. We start with a theoretical discussion of the
generalized Weyl Hamiltonian in Sec. II. Sec. III is devoted to the LL formation
in a strong magnetic field, for the case of tilted Dirac cones. Possible experimental
realizations in distorted graphene and \BEDT~are discussed in Sec. IV, which we
conclude with an analysis of a possible quantum Hall effect in \BEDT.

\section{Generalized Weyl Hamiltonian}

We consider a model of two-spinor fermions restricted to a 2D space.
Whereas the two-spinor form is in general dictated by relativistic
invariance in two space dimensions, it naturally arises in the
condensed matter situation of a lattice with two inequivalent sites.
The most general Hamiltonian linear in the 2D wave vector
$\bk=(k_x,k_y)$, is given by the ``generalized Weyl Hamiltonian",
\beq\label{WeylH} H=\sum_{\mu=0,...,3}\bv_{\mu}\cdot\bk \,
\sigma^{\mu}, \eeq in terms of the velocities
$\bv_{\mu}=(v_{\mu}^x,v_{\mu}^y)$, and the $2\times 2$ Pauli
matrices $\sigma^0\equiv \bone,
\vec{\sigma}=(\sigma^1,\sigma^2,\sigma^3)$. Here and in the
following parts, we choose a unit system with $\hbar\equiv 1$. Both
2D space components of the velocities,
$v_{\mu}^x=(v_0^x,\vec{v}^x)\equiv (v_0^x,v_1^x,v_2^x,v_3^x)$ and
$v_{\mu}^y=(v_0^y,\vec{v}^y)\equiv (v_0^y,v_1^y,v_2^y,v_3^y)$, are
in themselves vectors in the 4D spin space [the space of SU(2)
matrices] spanned by the Pauli matrices. The usual 2D Weyl
Hamiltonian, which describes for instance low-energy massless
electrons in graphene,\cite{antonioRev} is included in (\ref{WeylH})
if one considers $\bv_0=\bv_4=0$, $\bv_1=(v_F,0)$, and
$\bv_2=(0,v_F)$, in terms of the Fermi velocity $v_F$.

Although, at first sight, the Weyl Hamiltonian is described by eight
different parameters, given by
the four two-component velocities $\bv_{\mu}$, it is indeed overspecified. In order
to illustrate this point, we rewrite the Hamiltonian (\ref{WeylH}) in a different
manner,
\beq\label{WeylH2}
H=\bv_0\cdot\bk\, \sigma^0+\left(\vec{v}^x k_x+\vec{v}^y k_y\right)\cdot\vec{\sigma}.
\eeq
One may get rid of two parameters
($\bv_3=0$) by choosing the 3-quantization axis in the SU(2) space perpendicular to
the vectors $\vec{v}^x$ and $\vec{v}^y$.

This point is indeed remarkable and needs to be discussed in the light of graphene
physics. In this case, a constant $\sigma^3$ term breaks the inversion
symmetry of the
honeycomb lattice, e.g. due to a different on-site energy of
the two triangular sublattices. Usually, this gives rise to a mass term and breaks
the particle-hole symmetry. In the generalized Weyl Hamiltonian, this is not the
case because the $\sigma^3$ term is linear in the wave vector and therefore does
not affect the zero-energy state at $\bk=0$.

One may furthermore reduce the number of relevant model parameters by a simple
rotation of the 2D frame of reference, accompanied by a unitary transformation
in the SU(2) space, which leaves the 3-quantization axis invariant. One, thus,
obtains the ``minimal'' Weyl Hamiltonian
\beq\label{WeylMin}
H=\bw_0\cdot\bq\, \sigma^0+w_xq_x\sigma^x+w_yq_y\sigma^y,
\eeq
in terms of the four effective velocities $\bw_0=(w_{0x},w_{0y})$, $w_x$ and
$w_y$. A detailed discussion of the involved transformations and a derivation
of the exact expressions for the effective velocities may be found in the
Appendix \ref{app}.

The diagonalization of the minimal Weyl Hamiltonian yields the energy dispersions
\beq\label{Endisp}
\epsilon_{\lambda}(\bq)=\bw_0\cdot\bq+\lambda\sqrt{w_x^2q_x^2+w_y^2q_y^2},
\eeq
where $\lambda=\pm$ plays the role of the band index.

\begin{figure}
\epsfysize+6.0cm
\epsffile{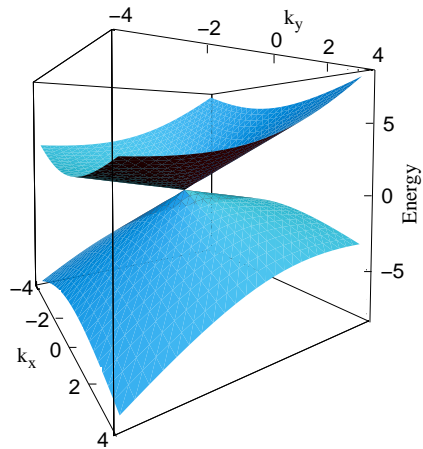}
\caption{Energy dispersion (\ref{Endisp}) for the special choice of $w_x=w_y=1$,
and $\bw_0=(0,0.6)$, in natural units. The Dirac cone is tilted in the
$y$-direction.}
\label{fig01}
\end{figure}

For $w_x=w_y=v_F$ and $\bw_0=0$, one obtains the isotropic
model, which applies e.g. to the low-energy electronic properties in graphene: the
Fermi velocities are the same in the $x$- and $y$-direction. The rotational symmetry
is broken if $w_x\neq w_y$ (anisotropic model). Such case may be obtained e.g. if
the graphene sheet is constrained by a uniaxial pressure, as is discussed in Sec.
IV A. For $\bw_0\neq 0$, the Dirac cones are tilted away from the $z$-axis,
as is shown in Fig. \ref{fig01}.

Notice that not all values of the tilt parameter $\bw_0$ are indeed physical. In
order to be able to associate $\lambda=+$ to a positive and $\lambda=-$ to a
negative energy state, one obtains the condition
\beq\label{TiltCond}
\left(\frac{w_{0x}}{w_x}\right)^2+\left(\frac{w_{0y}}{w_y}\right)^2<1.
\eeq
Unless this condition is satisfied, the iso-energetic lines are no longer ellipses
but hyperbolas. Notice that, here, we aim to use the generalized Weyl Hamiltonian
(\ref{WeylH}) and its resulting energy dispersion (\ref{Endisp}) to describe the
electronic properties of particular 2D materials. Although it may be interesting
to speculate about the resulting properties of a model that violates the condition
(\ref{TiltCond}), we are not aware of any physical example which might correspond
to such a case.

In a 2D lattice system with valley degeneracy, a generalized Weyl Hamiltonian
may describe the low-energy excitations in different valleys separately. In the
remainder of this paper, we will in general only consider a single valley (and
explicitly mention the inclusion of the twofold valley degeneracy
when needed). Note also that we do not consider the true electron
spin and do not include the corresponding twofold spin degeneracy.

In order to discuss the symmetry properties of the generalized Weyl Hamiltonian
(\ref{WeylMin}), it is convenient to introduce the unitary and Hermitian chirality
operator
\beq\label{chiral}
\mathcal{C}=\frac{w_xq_x\sigma^x+w_yq_y\sigma^y}{\sqrt{w_x^2q_x^2+w_y^2q_y^2}},
\eeq
which commutes naturally with the Hamiltonian. The associated eigenvalues
are $\alpha=\pm 1$ and coincide with the band indices $\alpha=\lambda$.
As exemplified in Sec. IV, this is
generally not the case in a physical condensed-matter situation --
the Weyl Hamiltonian corresponds to the effective model
at Dirac points, where the conduction band touches the valence band; these Dirac
points occur in pairs, at inequivalent points in the first BZ, which yields a
twofold valley degeneracy. In this case, the effective model is rather given by
$\xi H$, where $\xi=\pm$ denotes the two valleys, and the relation between
band index, chirality, and valley index is given by
\beq\label{BCV}
\lambda=\xi\alpha.
\eeq
In the present discussion, we may however identify the band index with the
chirality, for simplicity.

The eigenstates of the chirality operator are
\beq\label{ES}
\Psi_{\alpha}=\frac{1}{\sqrt{2}}\left(
\begin{array}{c}
e^{-i\phi_{\bk}} \\ \alpha
\end{array}\right)\ ,
\eeq
where $\tan\phi_{\bk}\equiv w_yk_y/w_xk_x$. These eigenstates are also the
natural eigenstates for the generalized Weyl Hamiltonian.

\section{Tilted Dirac Cones in a Magnetic Field}
We use the Peierls substitution to obtain the generalized Weyl Hamiltonian
in a magnetic field
\beq\label{PeierlsSub}
\bq\rightarrow \Pib=\bq+e\bA,
\eeq
where $\bA$ is the vector potential that generates the (uniform) magnetic field
$B\be_{z}=\nabla\times\bA$ perpendicular to the 2D plane. With the help of the ladder
operators
\beqn\label{ladderOp}
\nn
a &=& \frac{l_B}{\sqrt{2w_xw_y}}\left(w_x\Pi_x-iw_y\Pi_y\right),\\
a^{\dagger} &=& \frac{l_B}{\sqrt{2w_xw_y}}\left(w_x\Pi_x+iw_y\Pi_y\right),
\eeqn
in terms of the magnetic length $l_B=1/\sqrt{eB}$, one obtains the Hamiltonian
\beq\label{WeylB}
H_B=\frac{\sqrt{ 2 w_xw_y}}{l_B}\left(\begin{array}{cc}
 \frac{\wtilde_0}{2}(a e^{i \varphi}+a^{\dagger}e^{-i \varphi} ) & a \\
 a^{\dagger} & \frac{\wtilde_0}{2}(a e^{i \varphi}+a^{\dagger}e^{-i \varphi} )
\end{array}
\right).
\eeq
where we have defined
$$\wtilde_0 e^{i\varphi}\equiv \frac{w_{0x}}{w_x}+i\frac{w_{0y}}{w_y} ,$$
in terms of the effective tilt parameter 
\beq\label{tiltparam}
\wtilde_0\equiv\sqrt{\left(\frac{w_{0x}}{w_x}\right)^2+
\left(\frac{w_{0y}}{w_y}\right)^2}. 
\eeq

Instead of the full solution of the Hamiltonian
(\ref{WeylB}), we consider the effect of the magnetic field in a
semiclassical treatment. The Onsager relation\cite{onsager} states
that the surface $S(\epsilon)$ enclosed by a trajectory of constant
energy $\epsilon$ in reciprocal space is quantized as
$$S(\epsilon)l_B^2=(2\pi)^2\int_0^{\epsilon}d\epsilon'\, \rho(\epsilon')=2\pi(n+\gamma),
$$ where $n$ is an integer denoting the energy level which coincides
with the Landau level in the full quantum treatment. The additional
contribution $\gamma$ is related to a Berry phase acquired by an
electron during its cyclotron orbit. Usually, one has $\gamma=1/2$
except if there is an extra Berry phase of $\pi$, which in our case
yields $\gamma=0$, as in the case of graphene with no
tilt.\cite{mikitik} If one considers a density of states which
scales as $\rho(\epsilon)\propto \epsilon^{\alpha}$, the energy
levels thus scale as \beq\label{scaling} \epsilon_n\sim
[B(n+\gamma)]^{1/(1+\alpha)}, \eeq in the large-$n$ limit. In usual
(non-relativistic) 2D electron systems, one finds a constant density
of states, i.e. $\alpha=0$, and $\gamma=1/2$. The scaling of the
conventional Landau levels is therefore $\epsilon_n\propto
B(n+1/2)$. In the relativistic case of electrons in graphene, the
density of states vanishes linearly at the Dirac points, and one
therefore obtains $\epsilon_n\propto \sqrt{Bn}$ because $\alpha=1$
and $\gamma=0$. The relation (\ref{scaling}) has been generalized to
the case of a spatially anisotropic density of states  by Dietl {\sl
et al.} \cite{dietl}

From the scaling argument (\ref{scaling}) in the large-$n$ limit,
one may notice that the $B$-field scaling of the levels must be the
same as the $n$ scaling. Furthermore, one sees from the quantum
Hamiltonian (\ref{WeylB}) that the energy must scale as
$1/l_B\propto\sqrt{B}$. Therefore, the energy levels must obey, in
the large-$n$ limit, the equation \beq\label{LLraw}
\epsilon_{\lambda,n}\simeq\lambda\sqrt{2}\frac{v_F^*}{l_B}\sqrt{n},
\eeq as in the case of the Weyl equation for massless charged
particles, such as in graphene, apart from a renormalization of the
Fermi velocity.

The renormalization of the Fermi velocity may be obtained from the
calculation of the density of states. The total number of states
below a given energy $\epsilon$ within the positive energy cone is
given by \beqn \nn
N_+(\epsilon)&=&\frac{1}{(2\pi)^2w_xw_y}\int_{\epsilon_+(\qtilde)\leq\epsilon}
d\qtilde_x d\qtilde_y\\
\nn &=&\frac{1}{2\pi v_F^{*2}}\frac{\epsilon^2}{2}, \eeqn where we
have defined $\qtilde_{x/y}\equiv w_{x/y}q_{x/y}$, and the
renormalized Fermi velocity is written in integral form,
\beq\label{FermiVel}
\frac{1}{v_F^{*2}}=\frac{1}{w_xw_y}\int_0^{2\pi}\frac{d\phi}{2\pi}\frac{1}{
(1+\wtilde_0\cos\phi)^2}. \eeq in terms of the effective tilt
parameter (\ref{tiltparam}). One notices from Eq.
(\ref{FermiVel}) that if the condition (\ref{TiltCond}),
$|\wtilde_0|^2<1$, is not satisfied, the expression under the
integral diverges because the denominator may become zero. This
result is not suprising because the Onsager quantization relation,
which yields the energy levels (\ref{LLraw}) is only valid for
closed orbits, given e.g. by the elliptic isoenergetic lines. As
already mentioned, the orbits for $|\wtilde_0|\geq 1$ are open
hyperbolas, and the expression (\ref{LLraw}) is no longer valid.

\begin{figure}
\epsfysize+5.0cm
\epsffile{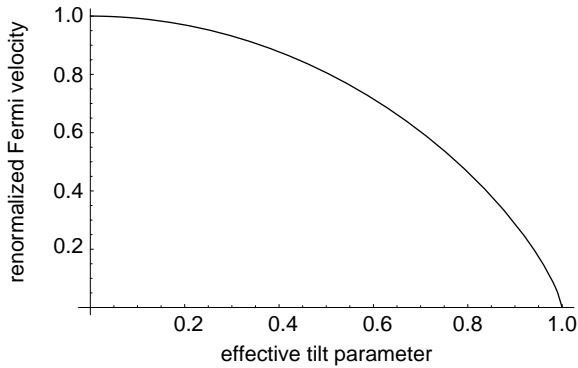}
\caption{Renormalized Fermi velocity $v_F^*/\sqrt{w_xw_y}$ as a function of
the effective tilt parameter $\wtilde_0\equiv\sqrt{(w_{0x}/w_x)^2+(w_{0y}/w_y)^2}$.
The Fermi velocity vanishes for $\wtilde=1$, where the orbits change from
ellipses to hyperbolas.}
\label{fig02}
\end{figure}

The density of states is obtained by differentiation of the number
of states, \beq\label{DOS} \rho(\epsilon)=\frac{|\epsilon|}{2\pi
v_F^{*2}}, \eeq which is the concise expression for both the
positive and negative parts of the tilted Dirac cones.

The $\sqrt{nB}$ behavior of Eq. (\ref{LLraw}) is, strictly speaking,
valid only in the large-$n$ limit. However, usually it yields
extremely good estimates for the levels down to values as small as
$n=1$. Special care is needed for the discussion of the $n=0$ level,
which requires a quantum treatment of the Hamiltonian (\ref{WeylB}).
In the following, we discuss the fate of the zero-energy
Landau level.

The behavior of this level may be understood with the help of the
quantum treatment of the Hamiltonian for $\bw_0=0$. In this case, the
expression (\ref{LLraw}) is exact with $v_F^*=\sqrt{w_xw_y}$, which
is also the $\bw_0=0$-limit of the expression (\ref{FermiVel}).
There exists thus a zero-energy level for $n=0$, which has the same
degeneracy, $N_B$ as all other levels $(\lambda,n)$, in terms of the
number of flux quanta $N_B=AB/(h/e)$ threading the total surface
$A$.

For non-zero values of $\bw_0$, the Hamiltonian
(\ref{WeylB}) may not be diagonalized by a simple canonical
transformation. However, the Hamiltonian (\ref{WeylB}) is
transformed as $H_B\rightarrow -H_B$ under space inversion,
$\br\rightarrow -\br$, as shown in the Appendix \ref{appB}. This
implies that the energy spectrum is symmetric around zero energy
(see Appendix \ref{appB}). Therefore, starting from $\bw_0=0$ and
adiabatically turning on $\bw_0\neq 0$, there are only two
possibilities for the evolution of the zero-energy level: (i) either
it remains at zero energy or (ii) it splits into (at least) two
sublevels $0^+$ and $0^-$, which are symmetric around zero energy.
However, splitting of the zero-energy level into sublevels can be
excluded on account of the degeneracy of this level. Indeed, when
$\bw_0=0$, the exact degeneracy of the zero energy ($n=0$) Landau
level is given by $N_B$ (remember that we only consider a single
valley here). When $\bw_0\neq 0$, it can, therefore, not split since
this would indeed lead to an unphysical doubling of the number of
quantum states because each level, $0^+$ and $0^-$, would have to be
$N_B$ times degenerate. Therefore, for all magnetic field strength,
there exist a zero-energy Landau level. The explicit expressions
for the zero-energy wave functions may be found in the Appendix \ref{appC}.
Notice, that this is
consistent with the semiclassical spectrum with $\gamma=0$.

In the above treatment, we only considered a single valley. We note
however that the magnetic field might introduce a coupling between
the two valleys. In such a case, we do not exclude a parity anomaly
which consists of a different behavior of the $n=0$ level at the two
inequivalent Dirac points at non-zero wave vectors in a lattice
model. In this case, space inversion would involve the low-energy
Hamiltonians at both Dirac points, and the spectrum is only
symmetric around zero energy if one accounts for both valleys. The
parity anomaly is, however, expected to play no physical role in the
continuum limit with $a/l_B\rightarrow 0$, where $a$ is the lattice
spacing.

In conclusion, we have obtained the semiclassical spectrum of Landau
levels [see Eqs (\ref{LLraw})-(\ref{tiltparam}], valid when $n\gg
1$) and checked that the zero energy level ($n=0$) indeed exists in
a full quantum treatment. Based on this two calculations, we expect
the semiclassical spectrum
to be a very good approximation to the true quantum spectrum of Landau
levels, for all $n$. This is one of the main result of the present
paper.

\section{Physical Examples of Tilted Dirac Cones}

After this rather technical discussion of the generalized Weyl Hamiltonian and
tilted Dirac cones, we discuss, here, two physical systems which may display
these properties. We find that whereas the tilt of the Dirac cones is well pronounced
and thus strongly affects the Landau level quantization in \BEDT, it is much
more difficult to induce a tilt in graphene via a lattice deformation. However,
a quinoid-type lattice deformation is also discussed for pedagogical reasons because
the general physical origin of tilted Dirac cones becomes transparent.

\subsection{Quinoid-type graphene under uniaxial strain}

\begin{figure}
\epsfysize+4.0cm
\epsffile{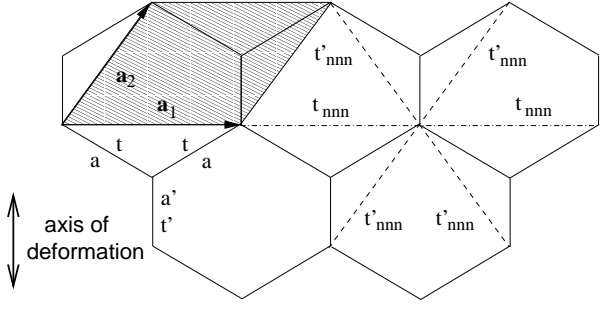}
\caption{Quinoid-type deformation of the honeycomb lattice -- the bonds parallel
to the deformation axis (double arrow) are modified. The shaded region indicates
the unit cell of the oblique lattice, spanned by the lattice vectors $\ba_1$ and
$\ba_2$. Dashed and dashed-dotted lines indicate next-nearest neigbors, with
characteristic hopping integrals $t_{nnn}$ and $t_{nnn}'$, respectively, which
are different due to the lattice deformation.}
\label{fig03}
\end{figure}

As a first example, we consider a graphene sheet which is deformed in one of its
principle symmetry axes. This particular deformation results in a quinoid variety
of the honeycomb lattice.\cite{saito} We treat its electronic properties within
the tight-binding approximation. Starting from the graphene honeycomb lattice,
with equal bond length $a\simeq 0.14$ nm and equal {\sl nn} hopping
energy $t\simeq 3$ eV, the bond length and hopping energy are
modified in the deformation axis (see Fig. \ref{fig03}),
$$a\rightarrow a'=a+\delta a \qquad {\rm and} \qquad
t\rightarrow t'=t+\frac{\partial t}{\partial a} \delta a,$$ and kept
unchanged otherwise. We call $\varepsilon=\delta a/a$ the relative
strain. Here, we consider a moderate deformation, $|\varepsilon| \ll
1$, such that one may linearize the hopping energy around its
nondeformed value $t$, and $\partial t/\partial a\simeq -5$ eV/${\rm
\AA}$.\cite{saito,phonon} This value agrees with an evaluation based
on Harrison's law\cite{Harrison} according to which
$t=C\hbar^2/ma^2$, where $C$ is a numerical prefactor of order one.
Derivation with respect to $a$ yields \beq\label{harrison}
\frac{\partial t}{\partial a}=-\frac{2t}{a}\sim -4.3\, {\rm eV/
\AA}. \eeq For simplicity and as a first approximation, one may keep
the bond angles fixed at $2\pi/3$. The underlying Bravais lattice is
no longer triangular but oblique with the basis vectors
$$\ba_1=\sqrt{3}a\be_x \qquad {\rm and} \qquad
\ba_2=\frac{\sqrt{3}}{2}a\be_x+\left(\frac{3}{2}a+\delta a\right)\be_y,$$
and the reciprocal lattice is spanned by the vectors
$$
\ba_1^*=2\pi\left(\frac{\be_x}{\sqrt{3}a}-\frac{\be_y}{3a+2\delta a}\right)\,\,\,\,
{\rm and} \,\,\,\,
\ba_2^*=\frac{4\pi\be_y}{3a+2\delta a}\ .
$$
Furthermore, we take into account {\sl nnn} hopping, with a
characteristic energy of\cite{nnnhopping} $t_{nnn}\simeq 0.1 t$ in the undeformed
horizontal axes. The deformation yields, in the same manner as for the {\sl nn} hopping
energies, different hopping energies for the other directions
(see Fig. \ref{fig03}),
$$t_{nnn}\rightarrow t_{nnn}'=t_{nnn}+\frac{\partial t_{nnn}}{\partial a}\delta a.
$$

\begin{figure}
\epsfysize+7.0cm
\epsffile{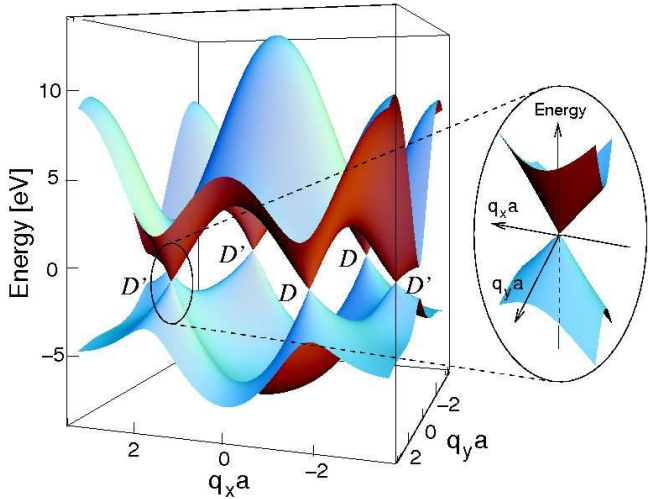}
\caption{Energy dispersion of the quinoid-type deformed the honeycomb lattice,
for a lattice distortion of $\delta a/a=0.4$, with $t=3$ eV, $t_{nnn}/t=0.1$,
$\partial t/\partial a=-5$ eV/${\rm \AA}$, and
$\partial t_{nnn}/\partial a=-0.7$ eV/${\rm \AA}$
The inset shows a zoom on one of the Dirac points, $D'$.
}
\label{fig03b}
\end{figure}

The tight-binding model may be described by the Hamiltonian
\beq\label{TBquinoid}
H=\sum_{\bq}\left(a_{\bq}^{\dagger},b_{\bq}^{\dagger}\right)\Hmath_{\bq}
\left(\begin{array}{c}
a_{\bq} \\ b_{\bq}
\end{array}
\right)
\eeq
in reciprocal space, where $a_{\bq}^{(\dagger)}$ and $b_{\bq}^{(\dagger)}$ are
the Fourier components of the annihilation (creation) operators on the
A and B sublattices, respectively. The Hamiltonian $2\times 2$ matrix
$$
\Hmath_{\bq}=\left(
\begin{array}{cc}
h'(\bq) & h^*(\bq) \\
h(\bq) & h'(\bq)
\end{array}
\right)
$$
is given in terms of the elements
\beqn\label{HTBoffD}
\nn
h(\bq) &=& -t\left[e^{i(q_y+\sqrt{3}q_x)a/2}+e^{i(q_y-\sqrt{3}q_x)a/2}
\right]\\
%\nn
&&
-t'e^{-iq_y(a+\delta a)}\\
\nn
&=& -2t\cos\frac{q_ya}{2}\cos\frac{\sqrt{3}q_xa}{2}
-t'\cos\left[q_y(a+\delta a)\right]\\
\nn
&&-i\left\{2t\sin\frac{q_ya}{2}\cos\frac{\sqrt{3}q_xa}{2}
-t'\sin\left[q_y(a+\delta a)\right]\right\}
\eeqn
and
\beqn\label{HTBD}
\nn
h'(\bq)&=& 2t_{nnn}\cos\sqrt{3}q_xa\\
\nn
&&+2t_{nnn}'
\left\{\cos\left[\frac{\sqrt{3}q_xa}{2}+
q_y\left(\frac{3}{2}a+\delta a\right)\right]\right.\\
&&\left.+\cos\left[-\frac{\sqrt{3}q_xa}{2}+
q_y\left(\frac{3}{2}a+\delta a\right)\right]\right\},
\eeqn
The energy dispersion is obtained from the eigenvalues of
$\Hmath_{\bq}$,
\beq\label{EnDispQuin}
\epsilon_{\lambda}(\bq)=h'(\bq)+\lambda |h(\bq)|
\eeq
and is plotted in Fig. \ref{fig03b} for a deformation of $\delta a/a=0.4$.
The two bands, $\lambda=+$ and $\lambda=-$, touch each other at the Dirac points
$\bq^D$, which are obtained from the condition $h(\bq^D)=0$,\cite{dietl}
\beq\label{DPquin}
q_y^D=0 \qquad {\rm and} \qquad q_x^Da=\xi \frac{2}{\sqrt{3}}\arccos\left(
-\frac{t'}{2t}\right),
\eeq
where $\xi=\pm$ denotes the two inequivalent Dirac points $D$ and $D'$,
respectively. In the
absence of any distortion, the Dirac points $D$ and $D'$ coincide with the
crystallographic points $K$ and $K'$, respectively, at the corners of the first BZ.
The distortion makes both pairs of points move in the same direction due to the
negative value of $\partial t/\partial a$. However, unless the parameters are
fine-tuned, this motion is different, and the two pairs of points no longer
coincide.\cite{DPmotion}

The low-energy properties of electrons in a quinoid-type distorted
graphene sheet are described by the linearized model around the Dirac points,
which is exactly of the form (\ref{WeylMin}) of the Weyl Hamiltonian,
\beq\label{WeylMinQuin}
H^{\xi}=\xi\left(\bw_{0}\cdot\bk\,\sigma^0+w_xk_x\sigma^x+w_yk_y\sigma^y\right),
\eeq
with the effective velocities
\beqn\label{VelQuin}
\nn
w_x &=& \sqrt{3}ta\sin\theta\\
w_y &=& \frac{3}{2}t'a\left(1+\frac{2\delta a}{3a}\right)\\
\nn
w_{0x} &=& 2\sqrt{3}\left(t_{nnn}a\sin 2\theta+t_{nnn}'a\sin \theta\right)\\
\nn
w_{0y} &=& 0,
\eeqn
where we have defined $\theta\equiv \arccos(-t'/2t)$. The corresponding energy
dispersion is independent of $\xi$, which is at the origin of the twofold valley
degeneracy. In order to obtain the concise form of Eq. (\ref{WeylMinQuin}),
we have chosen the spinor representation $(\psi_A,\psi_B)$ at the $\xi=+$
Dirac point and $(\psi_B,\psi_A)$ for $\xi=-$, i.e. interchanged the sublattice
components at $D'$. As mentioned in Sed. II, the relation between the band index
$\lambda$, chirality $\alpha$, and valley index $\xi$ is given by Eq. (\ref{BCV}),
$\lambda=\xi\alpha$,
due to the global sign $\xi$ in the Hamiltonian (\ref{WeylMinQuin}).
The constant term $h'(\bq=\xi\bq^D)\bone$ has been absorbed
in a renormalization of the chemical potential, the position of which is
determined by the electronic half-filling of the graphene sheet.

One notices from the Eqs. (\ref{VelQuin}) that the quinoid-type distortion
yields an anisotropy in the Fermi velocities, $w_x\neq w_y$, and that the
Dirac cones are tilted due to $w_{0x}\neq 0$. The isotropic graphene model is
retrieved at $\delta a=0$ -- one has then
$w_x=w_y=v_F=3ta/2\simeq 6.3$ eV\AA\ and $w_{0x}=w_{0y}=0$
because $t=t'$, $t_{nnn}=t_{nnn}'$, and $\sin \theta=\sqrt{3}/2=-\sin 2\theta$, in the undeformed case. Without deformation, {\sl nnn} hopping therefore
does not affect the energy dispersion at linear order, but only at second order.
This is due to the fact that the Dirac points are then situated at the
high-symmetry crystallographic points $K$ and $K'$. Indeed, this yields a parabolic
correction, which breaks the original electron-hole
symmetry.\cite{antonioRev,Peres}

To summarize, in order to obtain tilted Dirac
cones in graphene,
two ingredients are required: (i) {\sl nnn} hopping, which generates the
diagonal components $h'(\bq)$ in the Hamiltonian (\ref{TBquinoid}); and (ii)
for a linear contribution arising from this term, the Dirac points $D$
and $D'$ need to be shifted away from the high-symmetry points $K$ and $K'$. This
shift may be obtained by constraining the graphene sheet into such a quinoid type.

In the presence of a magnetic field, the LL spacing is affected by
the deformation because the Fermi velocity is renormalized according
to Eq. (\ref{FermiVel}), \beq\label{FermiVel2}
v_F^*\simeq\sqrt{w_xw_y}\left(1-\frac{3}{4}\wtilde_0^2\right), \eeq
for small values of the effective tilt parameter $\wtilde_0$. It may
be evaluated from the model parameters, \beqn\label{tiltQuin} \nn
\wtilde_0 &=& 2 \left(\frac{t_{nnn}}{t}\frac{\sin
2\theta}{\sin\theta}+\frac{t_{nnn}'}{t}
\right)\\
&\simeq& \frac{2}{t^2} (t t_{nnn}'-t' t_{nnn}). \eeqn In order to
estimate $t_{nnn}'$, we use the ``atomic orbitals overlap law''
familiar in the context of the extended H\"uckel model,\cite{salem}
$$
t_{nnn}(b,a) \approx t(a) e^{-(b-a)/d(a)}
$$
where $a$ is the {\sl nn} distance, $b$ is the {\sl nnn} distance,
and $d\approx a/3.5\approx 0.4$~\AA \, is a caracteristic
distance related to the overlap of atomic orbitals. In the
undeformed graphene $b=a\sqrt{3}$, whereas in the quinoid type
graphene $b'=b(1+\varepsilon/2)$ and $a'=a(1+\varepsilon)$. This
gives $t_{nnn}'=t_{nnn}(1-2\varepsilon +b\varepsilon/2d)$ and
$t'=t(1-2\varepsilon)$. Therefore, the effective tilt parameter is
given by
$$
\wtilde_0 \approx \frac{b}{d} \frac{t_{nnn}}{t}\varepsilon \approx
0.6 \varepsilon\, .
$$
As the correction to the Fermi velocity appears as
$1-3\wtilde_0^2/4$ [see Eq. (\ref{FermiVel2})], this effect remains
extremely small, and the tilt affects the LL spacing in a negligible
manner.

The main contribution to the renormalized Fermi velocity therefore
arises not from the tilt of the Dirac cones (effect of order
$\varepsilon^2$), but from the anisotropy in the Fermi velocities (effect
of order $\varepsilon$), and one finds \beq\label{FermVelQuin}
v_F^*\simeq v_F\left[1+\frac{1}{3}\left(\frac{\partial t}{\partial
a} \frac{\delta a}{t}+\frac{\delta a}{a}\right)\right]\simeq
v_F\left( 1-\frac{\varepsilon}{3}\right), \eeq which may yield an
experimentally observable effect in the percent range for a strain
of $\varepsilon \sim 10\%$.

>From an experimental point of view, such quinoid-type deformation
may be realized if one uses a piezoelectric substrate, on which the
graphene sheet is posed, instead of the most commonly used SiO$_2$.
Another possibility would be to use a mechanical deformation of the
underlying substrate. Such bending has been exploited e.g. to
investigate carbon nanotubes under strain.\cite{CNstrain} More
recently, graphene on polydimethylsiloxane (PDMS) has been put under
uniaxial strain by bending of the PDMS.\cite{huang} The elastic regime in
graphene requires that the strain is smaller than $10\%$ and the
rupture occurs around $20\%$. Therefore an upper bound for
$\varepsilon$ is certainly $10\%$.

\subsection{Organic 2D compounds}

Another example of a 2D metal, where tilted Dirac cones may occur, is the
layered organic compound \BEDT~under (uniaxial)
pressure.\cite{katayama,kobayashi,fukuyama} Each layer
may be described by an oblique lattice with four sites per unit cell, and the
electronic filling is $3/4$. In the vicinity of the Fermi energy, only two out of the
four bands are relevant for the low-energy electronic properties.
It has indeed been
shown that the band structure may be modeled with great precision within a
tight-binding model on a half-filled anisotropic triangular lattice with {\sl nn} and {\sl nnn}
hopping, where each site corresponds to a dimer.\cite{hotta}  This is a natural
assumption for $\kappa$- and $\lambda$-(BEDT-TTF)$_2$I$_3$, where
there exists one hopping energy which is largely enhanced with respect to the
others. In contrast to these compounds, the assumption may seem hasardous at first sight
in the case of \BEDT, where there is no such clearly enhanced hopping energy, such that
the dimerization is expected to be rather weak. Furthermore, these organic materials
exhibit strong electronic correlations, and a tight-binding calculation for quasi-free
electrons sweeps a lot of interesting physics under the carpet. However, the
high-pressure limit corresponds to a regime where the electrons are less strongly
correlated and where interaction effects may be taken into account via
renormalized effective hopping parameters.\cite{kobayashi}

\begin{figure}
\epsfysize+9.0cm
\epsffile{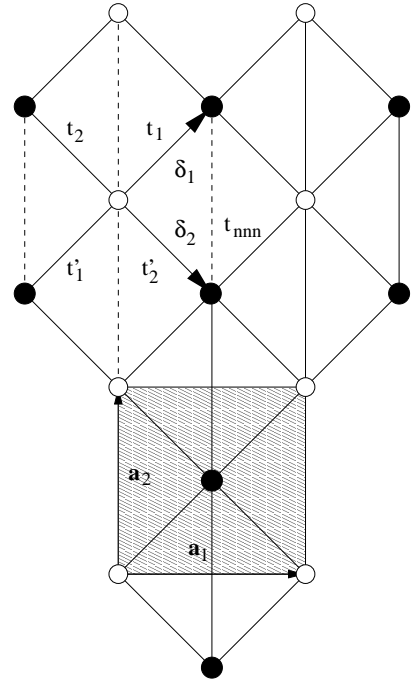}
\caption{Anisotropic triangular lattice model, with four different {\sl nn} hopping
energies, $t_1,t_1',t_2,$ and $t_2'$ and the {\sl nnn} hopping energy $t_{nnn}$. The unit
cell with two inequivalent sites is represented by the shaded region. The sites of
the $A$ and $B$ sublattices are depicted by the filled and open circles,
respectively. }
\label{fig04}
\end{figure}

The tight-binding model on the anisotropic triangular lattice is depicted in Fig.
\ref{fig04}.
The {\sl nn} are situated at the vectors $\pm \ba_1$ and $\pm \ba_2$, with
$$\deltab_1=\frac{1}{2}(\be_x+\be_y) \qquad {\rm and} \qquad
\deltab_2=\frac{1}{2}(\be_x-\be_y),$$
which connect sites on the different sublattices, $A$ and $B$ and
the vectors
$$
\ba_1=\deltab_1+\deltab_2=\be_x \qquad {\rm and} \qquad \ba_2=\deltab_1-\deltab_2=\be_y
$$
span the underlying Bravais lattice, which is chosen to be a square lattice, for simplicity.
Notice that the lattice may also be viewed as an anisotropic 2D NaCl lattice
(two inequivalent interpenetrating square lattices).
The bond length is set to unity, $a\equiv 1$. The {\sl nn} hopping energies
are $t_1$ and $t_1'$ in the directions $\pm\ba_1$, and $t_2$ and $t_2'$ in the
directions $\mp \ba_2$, respectively. The {\sl nnn} hopping energy is $t_{nnn}$.

The effective tight-binding model may be written in the same manner (\ref{TBquinoid})
as for the case of quinoid-type graphene, with the matrix elements
\beqn
\nn
h(\bq) &=& 2\left[(t_1+t_1')\cos\frac{q_x+q_y}{2}+(t_2+t_2')\cos\frac{q_x-q_y}{2}\right]
\\
\nn
&&+2i\left[(t_1-t_1')\sin\frac{q_x+q_y}{2}+(t_2-t_2')\sin\frac{q_x-q_y}{2}\right]
\eeqn
and
$$h'(\bq)=2t_{nnn}\cos q_y.
$$
The energy dispersion is obtained from Eq. (\ref{EnDispQuin}), and the
position of the Dirac points is calculated from
\beqn
\nn
\tan^2\frac{q_x^D}{2} &=& -\frac{\left(t_1'+t_2\right)^2-\left(t_1+t_2'\right)^2}{ \left(t_1'-t_2\right)^2-\left(t_1-t_2'\right)^2}\\
\nn
\tan^2\frac{q_y^D}{2} &=& -\frac{\left(t_1'+t_2'\right)^2-\left(t_1+t_2\right)^2}{ \left(t_1'-t_2'\right)^2-\left(t_1-t_2\right)^2}.
\eeqn
One may directly see that the r.h.s of both equations must be positive in order to
have a pair of Dirac points ($\bq^D$ and $-\bq^D$) within the first BZ,
$-\pi/2<q_x,q_y\leq \pi/2$.

An expansion around the Dirac points yields the generalized Weyl Hamiltonian
(\ref{WeylH}),
$$H^{\xi}=\xi\sum_{\mu=0}^2\bv_{\mu}\cdot\bk \sigma^{\mu},
$$
in terms of the velocities
\beqn\label{VelBEDT}
\nn
v_{0x} &=& 0,\qquad v_{0y}=-2t_{nnn}\sin q_y^D,\\
\nn
v_1^x &=& (t_1'+t_1)\sin\frac{q_x^D+q_y^D}{2}+(t_2'+t_2)\sin\frac{q_x^D-q_y^D}{2}\ ,\\
\nn
v_1^y &=& (t_1'+t_1)\sin\frac{q_x^D+q_y^D}{2}-(t_2'+t_2)\sin\frac{q_x^D-q_y^D}{2}\ ,\\
\nn
v_2^x &=& (t_1'-t_1)\cos\frac{q_x^D+q_y^D}{2}+(t_2'-t_2)\cos\frac{q_x^D-q_y^D}{2}\ ,\\
\nn
v_2^y &=& (t_1'-t_1)\cos\frac{q_x^D+q_y^D}{2}-(t_2'-t_2)\cos\frac{q_x^D-q_y^D}{2}\ .\\
\eeqn
Here, we have used the same spinor representation as for quinoid-type graphene, i.e.
we have interchanged the sublattice components when changing the valley.
One notices that the Dirac cones are tilted only if the Dirac points are not
situated at the border of the first BZ, $q_y^D=\pi/2$. This corresponds to the
high-symmetry crystallographic points in graphene, and {\sl nnn} hopping affects the
effective model again only at second order in the expansion around the Dirac points.

%%%%%%%%% new part
The experimental evidence for (tilted) Dirac cones in \BEDT~compounds under pressure
is yet rather weak. Whereas at ambiant pressure, the material is an insulator due
to charge ordering, temperature-dependent transport measurements under high hydrostatic
pressure have revealed a $T^2$ dependence of the carrier density below $50$ K,\cite{kajita,tajima}
as one would expect for relativistic electrons with a linear dispersion
relation.\cite{katayama,antonioRev} It is, however, not clear whether the compound has, under
these circumstances, a truely vanishing gap as for massless relativistic electrons
or whether a tiny gap persists. Furthermore the $T^2$ dependence of the carrier density is
accompanied by a temperature-dependent mobility, which results in an essentially constant
conductivity over a large temperature range.\cite{tajima}

A more direct evidence for the relevance of Dirac cones in \BEDT~
would be a measurement of the characteristic properties of relativistic
quantum Hall physics, as in the case of graphene.\cite{novoselov,zhang,antonioRev} The
following final part of this paper is devoted to the discussion of possible quantum
Hall physics in \BEDT.

%%%%%%%%%%% end new part

\subsection*{Possible quantum Hall effect in \BEDT}

Although it is a delicate issue to yield energy values for the hopping parameters
$t_1,t_1',t_2,t_2'$ and $t_{nnn}$ from the overlap integrals in
\BEDT,\cite{mori,kondo} we expect that the good agreement between
band-structure calculations in the full model with four sites per unit cell and
the anisotropic triangular lattice model\cite{hotta} yields the correct orders of
magnitude for the effective velocities (\ref{VelBEDT}). Using the prescription
proposed by Hotta\cite{hotta} and the overlap integrals calculated by Mori
{\sl et al.},\cite{mori} we may estimate $t_1=36$ meV, $t_1'=-86$ meV, $t_2=-24$ meV,
$t_2'=-77$ meV, and $t_{nnn}=-60$ meV. These values yield a pair of Dirac points
at $\bq^D$ and $-\bq^D$, with $\bq^D=(2.52,-3.08)$, in units of the inverse lattice
constant, which is on the order of $10$ \AA.\cite{mori,kondo,soderholm}
With the help of Eqs. (\ref{VelBEDT}), one thus obtains the effective velocities
$v_1^x=-0.035$ eV\AA, $v_1^y=0.315$ eV\AA, $v_2^x=-0.222$ eV\AA, $v_2^y=-2.121$ eV\AA,
$v_{0x}=0$, and $v_{0y}=0.074$ eV\AA. One notices a variation by almost two orders
of magnitude, and one may therefore expect rather large anisotropies.

The effective velocities in the minimal model are calculated with the help of
Eqs. (\ref{EffVel}), and one finds a rotation angle of $\theta=0.102$ and
the velocities $w_x=2.14$ eV\AA, $w_y=0.22$ eV\AA, $w_{0x}=-0.0075$ eV\AA, and
$w_{0y}=0.736$ eV\AA. The average Fermi velocity is therefore
$\sqrt{w_xw_y}=0.69$ eV\AA, which is roughly one order of magnitude smaller than
that in graphene. The tilt parameter (\ref{tiltparam}) is
$$\wtilde_0=0.33$$
and thus much larger than in the case of a quinoid-type deformation of
a graphene sheet. The tilt therefore leads to a reduction of the average Fermi
velocity, and one finds from Eq. (\ref{FermiVel2}) a renormalized velocity of
$$v_F^*\simeq 0.92 \sqrt{w_xw_y}\simeq 0.63 {\rm eV\AA}.$$

The renormalized Fermi velocity allows one to extract the typical energy scale for
the Landau levels in \BEDT, and one finds from Eq. (\ref{LLraw})
$\epsilon_{\lambda,n}=\lambda\omega_C^* \sqrt{n}$, with a characteristic ``cyclotron''
frequency of
\beq\label{cyclBEDT}
\omega_C^*=\sqrt{2}\frac{v_F^*}{l_B}\simeq 3.4 \sqrt{B[T]}\, {\rm meV},
\eeq
which is, due to the smaller Fermi velocity, roughly one order of magnitude smaller
than that in graphene. However, this energy scale is comparable to the cyclotron
frequency in GaAs heterostructures ($\omega_C\simeq 1.6 B[T]$ meV), which are
most commonly used in the study of quantum Hall physics.\cite{perspectives} One
may therefore expect that a relativistic quantum Hall effect\cite{novoselov,zhang}
could principally also occur in \BEDT\ if disorder does not prevent LL formation.

Experimentally, thin (BEDT-TTF)$_2$I$_3$ films have already been synthesized.\cite{laukhina}
Alternatively, one may hope that the
exfoliation technique,\cite{geim} which has proven to be particularly
successful in the fabrication of single-layer graphene sheets, also yields
reasonably thin \BEDT\ samples. However, (BEDT-TTF)$_2$I$_3$ crystals are
generally of lower mechanical stability than carbon crystals,
due to the relatively large lattice constants and the reduced binding
energies.

Apart from a direct measurement of a quantum Hall effect in \BEDT\ compounds, one
may probe the system via transmission spectroscopy in a magnetic field. This would
allow for a direct measurement of the cyclotron frequency and for a check of the relativistic character of electrons in \BEDT. Transmission spectroscopy
has indeed been
successfully applied to epitaxial\cite{sadowski} and exfoliated\cite{jiang}
graphene and yields a $\sqrt{B}(\sqrt{n+1}\pm \sqrt{n})$ scaling of the transmission
lines, as expected for the relativistic quantum Hall effect in graphene.

\section{Conclusions}

In conclusion, we have investigated tilted Dirac cones in deformed graphene and the organic
2D material \BEDT. The low-energy electronic properties are described by a generalized
Weyl Hamiltonian, which may in both physical systems be derived from a tight-binding model
on a lattice with two inequivalent sites. Whereas the presence of pairs of Dirac points is due
to {\sl nn} hopping, which couples neighboring sites on inequivalent sublattices, the tilt of the
Dirac cones arises from {\sl nnn} hopping if the Dirac points are shifted away from the points of
high crystallographic symmetry in the first Brillouin zone.

In the presence of a strong magnetic field, a semiclassical analysis yields
the same structure of relativistic LLs
as in non-deformed graphene, but with a renormalized effective Fermi velocity due to the tilt
and the anisotropy of
the Dirac cones. Whereas this effect is expected to be small in a quinoid-type deformation of the
graphene, our estimates for the effective velocities for \BEDT~indicate that the tilt yields a
significant reduction of the effective Fermi velocity, which determines the LL spacing. The largest
spacing of the $0\rightarrow +1$ and $-1\rightarrow 0$ LL transitions is on the order of
$3.4\sqrt{B[T]}$ meV, which is on the order of the (equidistant) LL spacing in GaAs heterostructures
most commonly used in quantum Hall effect measurements. Such measurements in \BEDT, as well as
LL spectroscopy, may be a possible experimental verification of the yet weakly corroborated
presence of Dirac cones in \BEDT.

\section*{Acknowledgments}

We acknowledge fruitful discussions with H\'el\`ene Bouchiat, Natasha Kirova, Claude Pasquier,
Jean-Paul Pouget, and Yoshikazu Suzumura. This work is partially supported by the Agence Nationale de la
Recherche under Grant No. ANR-06-NANO-019-03.

\appendix
\section{Derivation of the minimal Weyl Hamiltonian}
\label{app}

In order to reduce the number of effective parameters in the Weyl Hamiltonian (\ref{WeylH2}),
one rotates the 2D reference system in the physical space,
\beqn
\nn
k_x&=&\cos\vartheta\, q_x+\sin\vartheta\, q_y\\
\nn
k_y&=&-\sin\vartheta\, q_x +\cos\vartheta\, q_y,
\eeqn
accompanied by a unitary transformation in the SU(2) space,
$$U(\theta)=\cos\frac{\theta}{2}\,\bone +i\sin\frac{\theta}{2}\,\sigma^z,$$
which leaves the $3$-quantization axis invariant and describes a rotation in
the $xy$-plane in the SU($2$) spin space,
\beqn
\sigma^1&=&\cos\theta\, \sigma^x+\sin\theta\, \sigma^y\\
\sigma^2&=&-\sin\theta\, \sigma^x+\cos\theta\, \sigma^y.
\eeqn
If one chooses
$$\tan\theta=\frac{v_1^x\sin\vartheta+v_1^y\cos\vartheta}{v_2^x\sin\vartheta+
v_2^y\cos\vartheta}
$$
and
\beqn
\nn
\tan 2\vartheta &=& -\frac{2(v_1^xv_1^y+v_2^xv_2^y)}{(v_1^x)^2+(v_2^x)^2
-(v_1^y)^2-(v_2^y)^2}\\
\nn
&=&-\frac{2\vec{v}^x\cdot\vec{v}^y}{|\vec{v}^x|^2-|\vec{v}^y|^2},
\eeqn
one obtains the ``minimal'' Weyl Hamiltonian (\ref{WeylMin}).
In terms of the original velocities, the minimal set of effective
parameters (the velocities $\bw_0=(w_{0x},w_{0y})$, $w_x$ and $w_y$) reads
\begin{widetext}
\beqn\label{EffVel}
\nn
w_{0x} &=& v_0^x\cos\vartheta - v_0^y\sin\vartheta\ , \qquad
w_{0y} = v_0^x\sin\vartheta + v_0^y\cos\vartheta\ ,\\
%\nn
w_x^2 &=& \frac{(v_1^x)^2+(v_2^x)^2+(v_1^y)^2+(v_2^y)^2}{2}
+\sqrt{\left[\frac{(v_1^x)^2+(v_2^x)^2-(v_1^y)^2-(v_2^y)^2}{2}
\right]^2+\left(v_1^xv_1^y+v_2^xv_2^y\right)^2}\ ,\\
\nn
w_y^2 &=& \frac{(v_1^x)^2+(v_2^x)^2+(v_1^y)^2+(v_2^y)^2}{2}
-\sqrt{\left[\frac{(v_1^x)^2+(v_2^x)^2-(v_1^y)^2-(v_2^y)^2}{2}
\right]^2+\left(v_1^xv_1^y+v_2^xv_2^y\right)^2}\ .
\eeqn
\end{widetext}

\section{Transformation of the Weyl Hamiltonian under space inversion}
\label{appB}

The ingredients which intervene in the Peierls substitution
(\ref{PeierlsSub}), $\bq\rightarrow -i\nabla + e\bA(\br)$, are
vectors, which transform as $\bV\rightarrow -\bV$ under space
inversion. Because of this transformation property and because the
generalized Weyl Hamiltonian (\ref{WeylH}) is linear in the
momentum, we have \beq\label{SI} H_B(\br,\nabla)=-H_B(-\br,-\nabla),
\eeq in space representation, where $H_B$ is given in Eq.
(\ref{WeylB}). We now consider a solution of the Schr\"odinger
equation,
$$H_B(\br,\nabla)\psi_{\epsilon}(\br)=\epsilon \psi_{\epsilon}(\br),$$
of energy $\epsilon$. Due to the property (\ref{SI}) under space inversion,
one finds that
\beq\label{Bsym}
H_B(\br,\nabla)\psi_{\epsilon}(-\br)=-H_B(-\br,-\nabla)\psi_{\epsilon}(-\br)\\
= -\epsilon\psi_{\epsilon}(-\br). \eeq This means that
$\psi_{\epsilon}(-\br)$ is a solution of the Schr\"odinger equation
with energy $-\epsilon$, and to each energy $\epsilon$ in the upper
energy band there exists one at $-\epsilon$ in the lower one. The
resulting Landau level energy spectrum is, therefore, symmetric
around zero energy.

Notice, however, that this argument is only valid for the
generalized Weyl Hamiltonian if one neglects the underlying lattice
model for which the Weyl Hamiltonian describes the low-energy
properties.
As mentioned in the text, Dirac points occur in pairs at
non-zero wave vectors, and one thus obtains a twofold valley
degeneracy $\xi=\pm1$. In this case, there is a relative minus sign
between the two copies of the Weyl Hamiltonian $\xi H_B$. Space
inversion, therefore, becomes an exact symmetry of the model and
involves both valleys.
%Similarly, the property (\ref{SI}) is lost
%when one includes the spin ${\bf S}$, the degeneracy of which is
%lifted by the Zeeman term $\propto {\bf S}\cdot{\bf B}$ that is
%invariant under space inversion.

\section{Zero-energy Landau level for the case with tilt}
\label{appC}

In this appendix, we show that there exists a zero energy mode for the
Hamiltonian $H_B$ (\ref{WeylB}). We may represent the ladder operators 
$a$ and $a^{\dagger}$ as
$$a e^{i \varphi}=\partial_x+x, \qquad a^{\dagger} e^{-i \varphi}=-\partial_x+x\ ,$$
which act on states described as functions of the real variable $x$.
The latter is not necessarily identical to the $x$- or $x'$-variable
in the plane, but should rather be viewed as an auxiliary variable
that allows for a formal solution of the zero-energy problem.

The Hamiltonian (\ref{WeylB}), thus, reads  
\beq H_B=\frac{\sqrt{ 2
w_xw_y}}{l_B}\left(\begin{array}{cc}
\wtilde_0 x & (x + \partial_x) e^{-i \varphi}\\
(x - \partial_x) e^{i \varphi} &\wtilde_0 x
\end{array}
\right), \eeq 
and we may search the solutions of the zero-energy
eigenvalue equation, $H_B\psi_0(x)=0$, with the ansatz
$$\psi_0(x)=\left(\begin{array}{c}
 \alpha \\ \beta
\end{array}
\right) e^{-\gamma x^2/2}.
$$
This yields the system of linear equations
$$\begin{array}{ccccc}
 \wtilde_0 \, \alpha &+& e^{-i\varphi} (1-\gamma)\, \beta &=& 0\\
 e^{i\varphi}(1+\gamma)\, \alpha &+& \wtilde_0\, \beta &=& 0\ ,
\end{array}
$$
which has a non-zero solution if we choose
\beq\label{gamma}
\gamma=\sqrt{1-\wtilde_0^2}.
\eeq
and, thus,
$$\alpha = -\frac{\wtilde_0 e^{-i\varphi}}{1+\gamma}\beta\ .
$$
Notice that we only obtain bound states, i.e. states that decay 
at large values of $x$, for $\wtilde_0<1$
($\gamma>0$ and real), which corresponds to the maximal-tilt
condition (\ref{TiltCond}).

\end{document}